\documentclass[12pt]{article}

\usepackage[utf8]{inputenc}
\usepackage{graphicx}
\usepackage{float}
\usepackage{amsmath, amssymb}
\usepackage{bm}
\usepackage{physics}
\usepackage{siunitx}
\usepackage{mathrsfs}
\usepackage{nicefrac}
\usepackage{geometry}
\usepackage{hyperref}
\title{Fast Transport of  Trapped Ultracold Atoms Using Shortcuts-to-Adiabaticity by Counterdiabatic Driving}

\author{Denuwan Vithanage, Skyler Wright, Edith Luveina-Joseph, \\Christopher Larson and Edward Carlo Samson\\
\small Department of Physics, Miami University, Oxford, Ohio 45056, USA\\
}
\date{(Dated: June 19, 2023)}
\begin{document}
\maketitle

\begin{abstract}
We numerically study the fast spatial transport of a trapped Bose-Einstein condensate (BEC) using shortcuts-to-adiabaticity (STA) by counterdiabatic driving (CD).  The trapping potential and the required auxiliary potential were simulated as painted potentials.  We compared STA transport to transport that follows a constant-acceleration scheme (CA).  Experimentally feasible values of trap depth and atom number were used in the 2D Gross-Pitaevskii equation (GPE) simulations.  Different transport times, trap depths, and trap lengths were investigated.  In all simulations, there exists a minimum amount of time necessary for fast transport, which is consistent with previous results from quantum speed limit studies.
\end{abstract}


\section{Introduction}
Research on quantum technologies can be classified into three categories: quantum metrology and sensing \cite{Borde.1997, Cronin.2009, Schaff.2014}, quantum communication \cite{Gisin.2007,Duan.2002}, and quantum simulation and computing \cite{Bloch.2012,Blatt.2012,Cirac.1995,Ladd.2010}.  Each technology category has its own set of challenges for future development and for their transition from laboratory research to field-deployable applications.  Nevertheless, all of them have one common problem that has to be addressed: the need for a scalable technology that can manipulate and control quantum systems at a fast or high repetition rate, while maintaining quantum fidelity, i.e. the condition in which the unique quantum properties that allow for the significant gains in precision and computation power are not degraded or are not eliminated from the quantum system.

As an example, the power of quantum computing devices depends on their scalability;  it is estimated that anywhere between 1000 and more than 10000 physical qubits are required to make a single logical qubit that can be used for a meaningful computation \cite{Fowler.2012}.  Current schemes for qubit control are limited to computations that use only tens of physical qubits \cite{Kielpinski.2002}.  Hence, to manage a full-scale quantum computer, new technology that allows optimal control of large-scale quantum systems needs to be developed.  A proposed architecture for large-scale computing involves a large interconnected network of computing sites \cite{Kielpinski.2002}.  To perform complex calculations, the qubits in each site would have to be shuttled around the network from one site to another, so that the calculations would be done in sequential stages while still manipulating a few number of qubits in each site.  One requirement for this architecture is that the transfer time between sites should consume a small fraction of a clock cycle, so that calculation speeds would not be limited.  Unfortunately, fast transport can cause decoherence of the qubits, which affects the amount of error in the system \cite{Kielpinski.2002}.

On the other end of scalability are quantum inertial sensors.  There is a drive for the development of small portable sensors that have precision and sensitivity (or better) comparable to those of large-scale atom interferometers and gravimeters \cite{Boshier.2010}, which have already achieved the same level of accuracy of free-fall corner-cube optical interferometers \cite{Peters.2001}.  Unfortunately, the footprint of state-of-the-art atom interferometers is large.  To develop portable quantum-based inertial sensors, it was proposed that BECs in a waveguide interferometer would be a more suitable quantum system \cite{Hinds.2001}.   A BEC-based waveguide interferometer requires the BEC to be physically split into two phase-coherent condensates and translated spatially when making a single measurement.  Multiple measurements at high repetition rates would significantly improve precision for these kinds of devices.  The fast manipulation of a BEC (e.g. during splitting or transport) presents a problem as it can cause axial and breathing mode excitations in the condensate, leading to dephasing and leading to large errors in the measurements \cite{Shin.2004, Collins.2005}.  Hence, similar to the case in quantum computing, new technology that can control and manipulate BECs at a fast rate while maintaining quantum fidelity is needed, but currently it is lacking.

Shortcuts to adiabaticity (STA) \cite{Chen.2010, Muga.2009, Torrontegui.2013,DelCampo.2019} are protocols that are designed to speed up an adiabatic process through a sequence of nonadiabatic steps, i.e., achieving the end-product of an infinitely slow adiabatic process in a very short time interval.  There are several types of STA protocols that have been studied theoretically: invariant-based inverse engineering \cite{Ibanez.2011,Torrontegui.2011,Muga.2011}, fast-forward approach \cite{Masuda.2010,Masuda.2011, Muga.2011}, shortcuts using unitary transformations \cite{Ibanez.2012}, shortcuts based on optimal control theory (\cite{Salamon.2009, Murphy.2009}, and transitionless quantum driving or counterdiabatic driving (CD) \cite{DelCampo.2012,DelCampo.2013,Deffner.2014}.  Among these different approaches, counterdiabatic driving allows for the engineering of STA in systems of arbitrary Hamiltonians, which makes it the most feasible shortcut scheme to be adaptable for any quantum device or application \cite{DelCampo.2019}.  

In CD, an auxiliary potential $V_{\text{CD}}\pqty{t}$ is designed and added to the ``original adiabatic evolution Hamiltonian $H_0\pqty{t}$ of the system so that the Hamiltonian driving the evolution of the system is now the sum of the two $\pqty\big{H\pqty{t} = H_0\pqty{t} + V_{\text{CD}}\pqty{t}}$.  The auxiliary term is designed to suppress the transitions between the different eigenstates of $H_0\pqty{t}$.  This enforces adiabatic following at the level of each eigenstate \cite{DelCampo.2013,DelCampo.2019}.  A recent theoretical work \cite{DelCampo.2013} further refined the CD approach by reformulating it in terms of scaling laws to the effect that it makes it easier to experimentally realize a CD potential using well-established experimental techniques (e.g. magnetic or optical trapping) for single-particle, many-body, and/or nonlinear systems in a variety of trapping potentials.  This makes it applicable to nonlinear systems such as a Bose-Einstein condensate and many-body systems like atom clounds in optical lattices \cite{DelCampo.2019}.

The structure of the designed auxiliary potential $V_{\text{CD}}\pqty{t}$ required by the CD approach may be different from the structure of the original Hamiltonian $H_0\pqty{t}$ and may lead to complex engineering in order to experimentally realize the STA scheme \cite{Torrontegui.2013}.  The painted potential technique is a method that can provide trapping potentials for Bose-Einstein condensates (BECs) that are simultaneously dynamic, completely arbitrary, and stable enough to not heat the BEC \cite{Henderson.2009}, which makes it ideal to generate $V_{\text{CD}}\pqty{t}$ for experiments.  Painted potentials are generated by a rapidly scanning red-detuned, focused laser beam that ``paints" a time-averaged optical potential on top of a static red-detuned light sheet (the ``canvas") \cite{Henderson.2009}.  The desired configuration and dynamics of the painted potentials can be easily specified by simply encoding the sequence of the trapping potential geometries to the control mechanism of the ``painting" beam.  Painted potentials have been used in several applications, such as the creation of matter wave Bessel beams and to induce quantized circulation in a BEC \cite{Ryu.2014}, the creation of Josephson junctions in a toroidal BEC as an atomic analog of a SQUID \cite{Ryu.2013}, and to realize a matter wave integrated circuit \cite{Ryu.2015}.

In this paper, we implement a counterdiabatic driving protocol to perform one-dimensional (1D) transport of a trapped Bose-Einstein condensate (BEC).  We model our system as being implemented in a device that uses time-averaged painted potentials \cite{Henderson.2009} to generate the potential trap and the auxiliary potential.  By solving the time-dependent Gross-Pitaevskii equation (GPE) numerically using a split-step Fourier method, we study different experimental parameters (transport time, trap depth, and painting beam width) and their effects on the post-transport quantum fidelity of the trapped BEC.

\section{Methods}
\subsection{Shortcut-to-adiabaticity protocol}\label{sec:STAprotocol}
References \cite{DelCampo.2013} and \cite{Deffner.2014} discuss the derivation of the necessary STA trajectory for any type of trapping potential, together with the instantaneous magnitude of the auxiliary potential.  In the following, we provide a short description of the STA protocol as it applies to the case of one-dimensional BEC transport.  

\begin{figure}[H]
\centering
\includegraphics[width=2.5in]{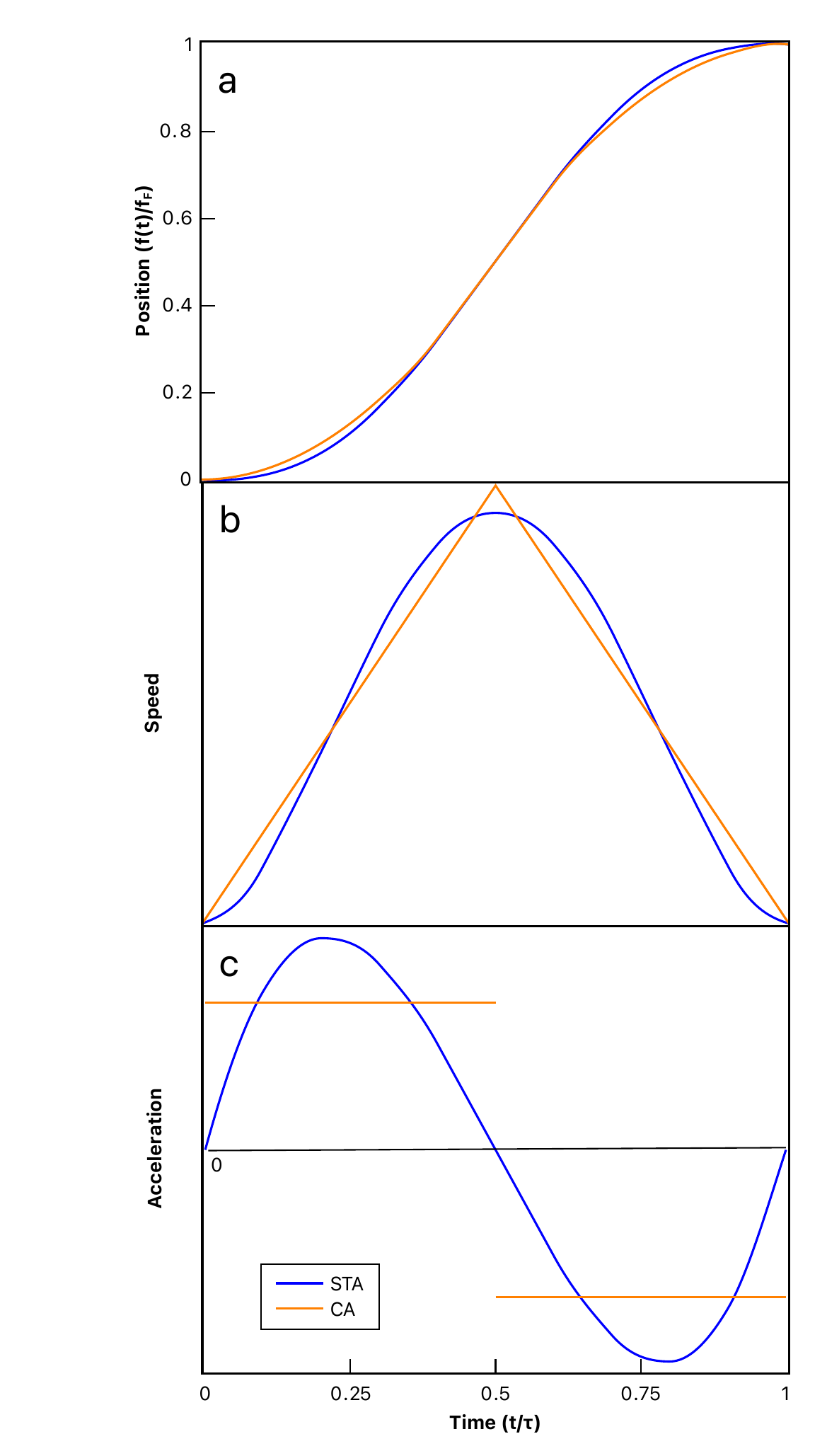}
\caption{A comparison of the (a) trajectories, (b) instantaneous speeds, and (a) accelerations between the STA protocol and a constant acceleration scheme. Note the relatively similar behaviors of the trajectories and speed profiles, but the STA acceleration follows the behavior of a third-order polynomial. }
\label{fig:STAtraj}
\end{figure}

\begin{figure}[H]
\centering
\includegraphics[width=4in]{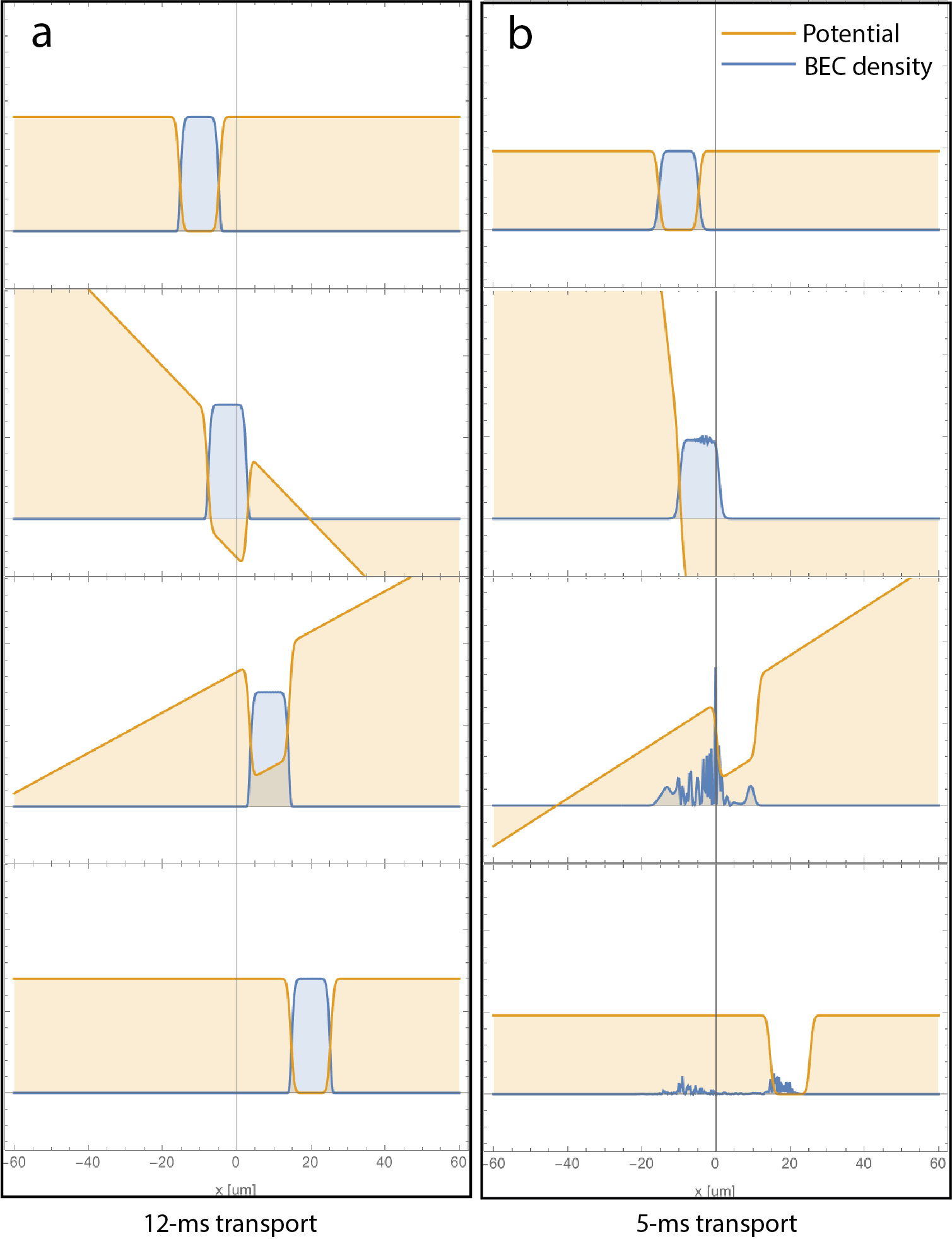}
\caption{Time evolution of the trapped BEC density (blue) and the trapping and auxiliary potentials (orange) during (a) a 12-ms and (b) a 5-ms transport. The 12-ms transport results into a quantum fidelity of 1, while the 5-ms transport has a low quantum fidelity, with atoms spilling out of the trap.}
\label{fig:BECevol}
\end{figure}

If the BEC needs to be transported from the initial position $\vb{f_0}$ to the final position $\vb{f_F}$, the trajectory $\vb{f}\pqty{t}$ of the center of the trapping potential is described by: 
\begin{equation}\label{eq:STA}
\vb{f}\pqty{t} = \vb{f_0} + \pqty{\vb{f_F}-\vb{f_0}}\bqty{10\pqty{\frac{t}{\tau}}^3 - 15\pqty{\frac{t}{\tau}}^4 +6\pqty{\frac{t}{\tau}}^5},
\end{equation}
where $\tau$ is the required transport time.  Compared to a trajectory using a constant-acceleration scheme (in which the potential trap is uniformly accelerated halfway through the trajectory and uniformly decelerated to the final position), the trajectory $\vb{f}\pqty{t}$ closely resembles and has only very small deviations from the CA trajectory (Figure \ref{fig:STAtraj}a).  The speed evolution of the STA protocol also follows the behavior of the CA scheme with small deviations (Figure \ref{fig:STAtraj}b). The main difference between the two protocols can be seen in the acceleration profile along the trajectory (Figure \ref{fig:STAtraj}c).  The STA acceleration constantly changes and follows a third-order polynomial behavior.
The instantaneous magnitude of the auxiliary potential $V_{\text{CD}}$ is given by:
\begin{equation}\label{eq:AuxPot}
V_{\text{CD}}\pqty{\vb{q},t} = -m\ddot{\vb{f}}\pqty{t}\vdot\vb{q},
\end{equation}
where $m$ is the atomic mass of \textsuperscript{87}Rb and $\vb{q}$ is a position vector measured with $\vb{f}\pqty{t}$.  $V_{\text{CD}}$ is a linear potential with a slope proportional to the instantaneous acceleration at any given time.  We note that at the start and end of transport, $V_{\text{CD}}$ is zero.  The evolution of the trapping potential and the effect of $V_{\text{CD}}$ during transport are shown in Figure \ref{fig:BECevol}.
\subsection{Numerical simulations}
We performed a two-dimensional (2D) split-step Fourier solution of the time-dependent Gross-Pitaevskii equation \cite{Pethick.2008}: 
\begin{equation}
i \hbar \pdv{t} \Psi \pqty{x,y;t} = \bqty{-\frac{\hbar^2}{2m}\laplacian_{x,y} + V_0 \pqty{x,y;t} + V_{\text{CD}} + g_{2\text{D}} \abs{\Psi \pqty{x,y;t}}^2}\Psi \pqty{x,y;t}
\label{eq:GPE}
\end{equation}
where $\Psi\pqty{x,y;t}$ is the BEC wavefunction and $m$ is the mass of a $^{87}\text{Rb}$ atom.  The 2D interaction parameter $g_{2\text{D}}$ is defined as: 
\begin{equation}
g_{2\text{D}} = \frac{4 \pi \hbar^2 a}{\sqrt{2\pi}m l_z}
\label{eq:g2d}
\end{equation}
where $a$ is the the atomic $s$-wave scattering length of the atoms and $l_z$ is an effective length parameter. Similar to a painted potential, the trapping potential $V\pqty{x,y}$ was modeled as a sum of $N$ two-dimensional Gaussian functions that are displaced by a distance $\Delta x$ from each other and spans the trap length:
\begin{equation}
V\pqty{x,y} = -U_0 \frac{\sum_{n = -N/2}^{N/2} \operatorname{exp} \bqty{-\frac{\pqty{x - x_0 - n\Delta x }^2 + y^2}{2 \sigma_r^2}}}{\sum_{n = -N/2}^{N/2} \operatorname{exp} \bqty{-\frac{\pqty{-x_0 - n\Delta x }^2}{2 \sigma_r^2}}},
\end{equation}
where $U_0$ is the magnitude of the trap depth, $\sigma_r$ is the beam waist of the painting beam and $\pqty{x_0,0} = \pqty{\vb{f}\pqty{t},0} $ is the trap center. For all simulations, the trap length is $\SI{10}{\micro\meter}$ and generated with $N = 25$.  The painting beam has $\sigma_r = \SI{1.5}{\micro\meter} $, unless otherwise noted.

A Bose-Einstein condensate consisting of 10000 atoms is held in a potential well that consists of a flat bottom and Gaussian-shaped walls.  The trap depth is $U_0 = k_B \cross \pqty{100,200,300,400,500}$ nK, where $k_B$ is Boltzmann's constant.  Simulations were performed using different transport distances $\vb{f}\pqty{\tau}$.  To evaluate the effectiveness of STA transport, the quantum fidelity $\mathscr{F}$ between the resulting BEC state after transport $\psi\pqty{\vb{r},\tau}$ and the BEC ground state of the trapping potential at the final location $\psi^{\text{GS}}_f\pqty{\vb{r}}$  is calculated:
\begin{equation}\label{eq:Fidelity}
\mathscr{F} = \abs{\braket{\Psi\pqty{\vb{r},\tau}}{\Psi^{\text{GS}}_f\pqty{\vb{r}}}}^2.
\end{equation}
\section{Discussion of Results}
\subsection{Transport Time}
\begin{figure}[H]
\centering
\includegraphics[width=5.2in]{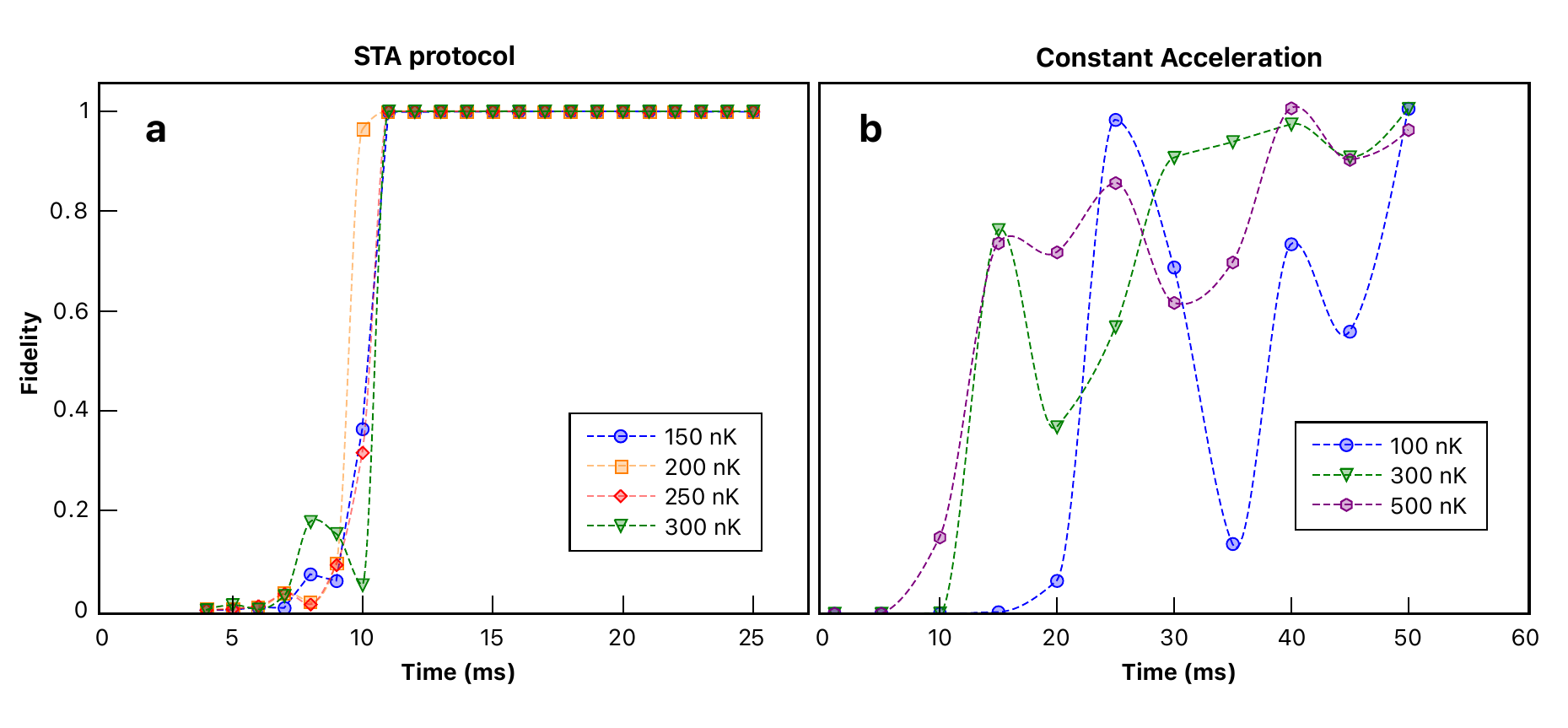}
\caption{The resulting quantum fidelities after transport of a BEC with $N = 10000 \text{ atoms}$ for (a) an STA protocol and (b) a constant-acceleration  scheme.}
\label{fig:STAtime}
\end{figure}

Figure \ref{fig:STAtime}a shows the post-transport fidelities at different transport times and different trap depths when the STA protocol was implemented.  At longer transport times, the trapped BEC has a quantum fidelity $\mathscr{F} \sim 1$, with $\tau \sim \SI{10}{\milli\second}$ as the fastest transport possible without decreasing the fidelity.  An example of the simulation in the high-fidelity region is shown in Figure \ref{fig:BECevol}a, which shows a transport time of $\SI{12}{\milli\second}$. As the potential well is translated, the BEC remains inside it while remaining in the ground state of the BEC.  Around this region, there is an abrupt transition to a region of transport times that results in $\mathscr{F} < 0.5$.  In comparison, there is also a region of high quantum fidelity in the constant-acceleration scheme (Figure \ref{fig:STAtime}b), albeit a fidelity of 1 is not obtained until transport times are greater than $\SI{20}{\milli\second}$.  The fidelities obtained at this region vary, and unlike in the STA scheme, it is still possible to obtain a lower fidelity at longer transport times, even when high fidelity has already been obtained at a faster transport.  
Based on the simulations, there are two possible mechanisms that can contribute to a decrease in fidelity after transport: atom loss and center-of-mass motion.  Atom loss refers to a decrease in the number of atoms inside the potential trap.  Center-of-mass (COM) motion is the oscillatory movement of atoms within the trap, which leads to a redistribution of the BEC atom density.  From the simulations of both transport schemes, atom loss is the major factor leading to a decrease in fidelity for transports in the region of low fidelity, while COM motion is for the region of high fidelity.  Atom loss is primarily due to atoms escaping from the fast-moving trap during transport, which effectively causes the atoms to ``spill out'' of the trap.  This can be seen in the simulations in Figure \ref{fig:BECevol}b, in which the atom density can be located outside the extent of the potential well. For transports having $\mathscr{F} \sim 1$, the primary mechanism of fidelity decrease is the presence of micro-oscillations of the BEC density distribution about the trap center along the long axis (direction of transport).

The presence of a distinct transition between the low-fidelity and high-fidelity regions in the STA simulations (Figure \ref{fig:STAtime}a) suggests that there is a maximum speed or a speed limit for BEC transport, even when using an STA protocol.  This can be interpreted as a consequence of the time-energy uncertainty principle: $\Delta t \gtrsim \hbar/\Delta E$. We can view BEC transport as a quantum state evolution from a Hamiltonian centered at an initial position $x_i$ to another Hamiltonian centered at a final position $x_f$.  The occurrence of the minimum transport times for the different trap depths in Figure \ref{fig:STAtime}a around the same region $\pqty{\sim \SI{10}{\milli\second}}$ can be interpreted as a consequence of this uncertainty relation \cite{Campbell.2017}.  Despite having different trap depths, the initial and final Hamiltonians for any of the transports are exactly identical except for the positions of the trap center.  Thus, for any trap depth, the energy transition would be around the same range of values, leading to similar minimum transition times.   If an STA transport protocol provides the optimal transition between displaced quantum states, then it has the possibility of being a testbed for quantum speed limit experiments.
\subsection{Trap Depth}
As shown in Figure \ref{fig:STAtime}, the STA simulations performed using different trap depths resulted in relatively similar values for the post-transport fidelity for a given transport time. This result is in line with the motivation to use an STA protocol, as described in Section \ref{sec:STAprotocol}, to transport trapped BECs, regardless of trap characteristics.  

In contrast, the fidelities from the CA transport scheme depend on the trap depth being used.  It can be noted that there are occurrences of a longer transport time having a lower fidelity than a faster transport, and then there would be an increase in fidelity again when the transport time is increased by a few milliseconds.  We conjecture that these decreases in fidelity occur at transport times that cause the trapped BEC to oscillate in resonance with the trap frequency along the transport direction.  The effective potential trap is anharmonic, but there is a characteristic trap frequency $\omega_{x}$ in which a BEC will have center-of-mass oscillations when displaced from the trap center.  To calculate the characteristic trap frequency, we performed numerical GPE simulations of a BEC that is initially displaced from the center of the potential trap.  We then observed the evolution of the BEC while in the nonmoving trap for $\SI{5}{\milli\second}$ and tracked the oscillation of its COM position, which yielded the characteristic trap frequency $\omega_x$.
\begin{figure}[H]
\centering
\includegraphics[width=3in]{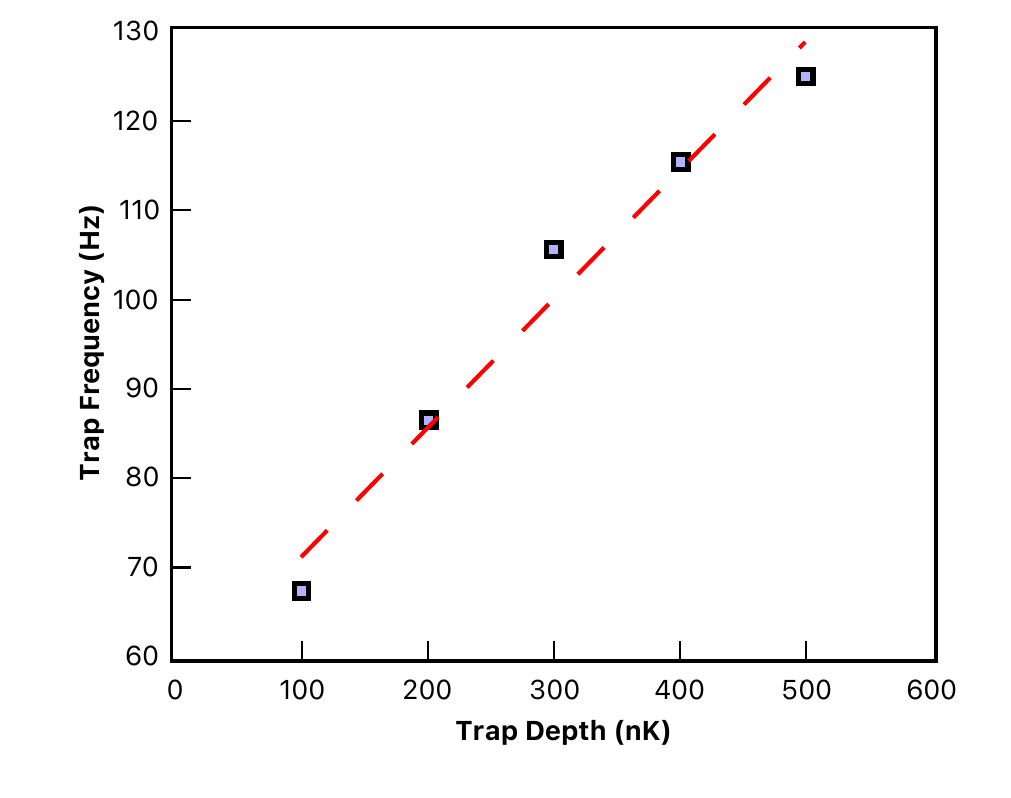}
\caption{The measured trap frequencies along the transport direction for a given trap depth.  The frequencies were calculated using GPE simulations of a trapped BEC that is initially displaced in a nonmoving trap and observing the COM oscillations for $\SI{5}{\milli\second}$.}
\label{fig:TrapFreq}
\end{figure}  
Shown in Figure \ref{fig:TrapFreq} are the measured $\omega_x$ from the GPE simulations. For the trap depth of $\SI{300}{\nano\kelvin}$, the characteristic frequency is $\sim \SI{106}{\hertz}$, giving a characteristic period of $\SI{9.4}{\milli\second}$.  We note that the first fidelity decrease for this trap depth occurs at $\tau = \SI{20}{\milli\second}$, which is approximately twice the characteristic trap period.  The same can be said for the other two trap depths presented in Figure \ref{fig:STAtime}b.  For the trap depth of $\SI{100}{\nano\kelvin}$, the characteristic trap period is $\SI{14.9}{\milli\second}$ and a decrease in fidelity occurs around $3\times$ and $4\times$ of this period $\pqty{\SI{29.8}{\milli\second} \text{ and } \SI{44.7}{\milli\second}\text{, respectively}}$.  

The CA scheme effectively is simple harmonic motion because the first half of the transport is at constant acceleration and the second half at a constant deceleration.  The effective time-averaged potential that moves the trap alone would be harmonic, and the whole trajectory in a CA scheme is the first half of an oscillation with the transport time as the oscillation period.  This interpretation is only valid if the trapped atoms move with the trap, which is the reason why in-resonance oscillations were not observed in shorter transport times $\pqty{< \SI{10}{\milli\second}}$.  Additionally, we note that the amount of decrease in fidelity decreases as the transport times become longer.  The CA scheme resulting into a harmonic potential can only happen at short time scales so that it can be viewed as time-averaged. 

\subsection{Beam width of painting beam}
\begin{figure}[H]
\centering
\includegraphics[width=3in]{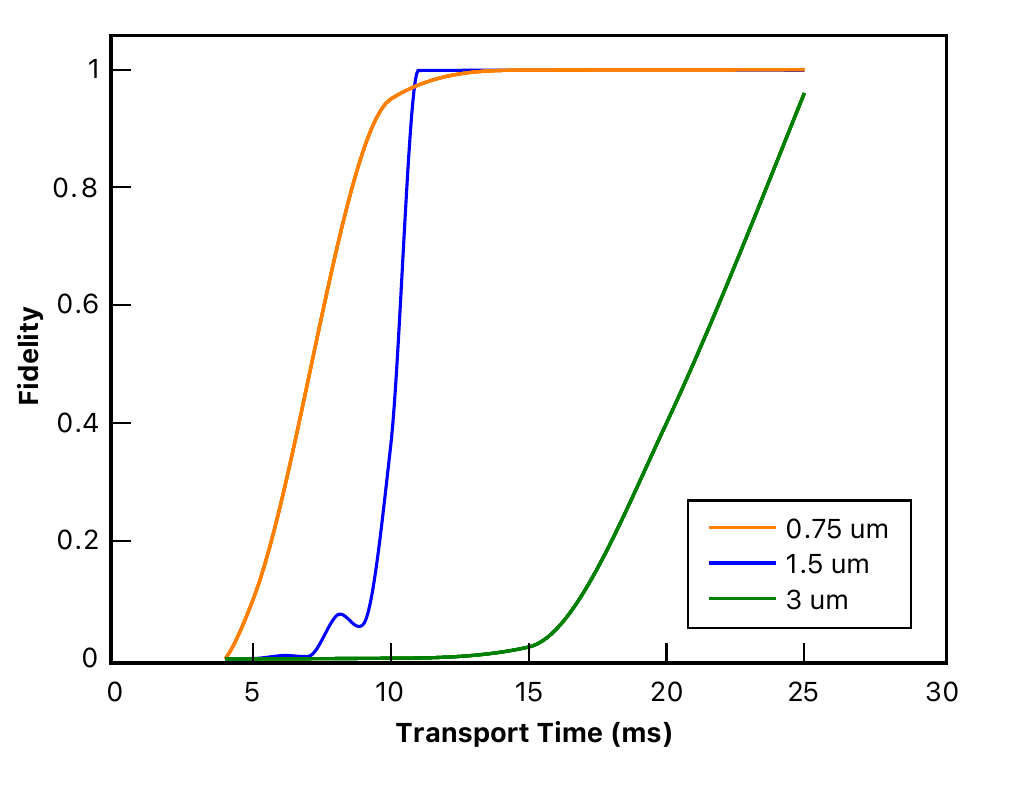}
\caption{Post-transport fidelities for different beam widths of the painting beam.  There is a shift in the location of the transition region between regions of low and high fidelities for different beam widths.}
\label{fig:BeamWidth}
\end{figure}

The STA transport simulations were also performed using different beam widths for the painting beam, and post-transport feadelities are shown in Figure \ref{fig:BeamWidth} when $U_0 = k_{\text{B}} \times \SI{150}{\nano\kelvin}$.  At each beam width, similar behaviors were observed for the post-transport fidelities: a region of low fidelity at short transport times and a region of high fidelity at long transport times with a transition in between.  As the beam width increases, the location of the transition changes to longer transport times. The shift in the transition region can be attributed to an increase in density oscillations during transport as the beam width increases.  Due to the larger painting beam, the trap frequency along the $y$-direction decreases and provides weaker confinement of the atoms, allowing for BEC density fluctuations.

\section{Conclusions}
We presented an analysis of the implementation of an STA protocol for transport of trapped Bose-Einstein condensates.  Using numerical simulations based on the Gross-Pitaevskii equation, the effects of different experimental parameters were studied.  Even with an STA protocol, there is a minimum transport time that can provide $\mathscr{F}\sim 1$.  This implies that these experiments can be used in future studies of quantum speed limits.  The effectivity of an STA protocol for BEC transport was shown when compared to a constant-acceleration scheme, as the resulting fidelity from an STA protocol is not affected by the trap frequency.  Lastly, the trap frequency along the direction perpendicular to the direction of motion can affect the resulting fidelity for a given transport time.  This suggests that further studies using three-dimensional simulations may provide additional insights for experimental implementation, as the trap frequency along the vertical direction can also affect the post-transport fidelity.

\vspace{6pt}

\subsubsection*{Author Contributions}
Conceptualization, E.C.S.; methodology, D.V., S.W., E.L.-J., C.L., and E.C.S.; software, D.V., S.W., and E.C.S.;  formal analysis, D.V., S.W., E.L.-J., C.L., and E.C.S.; manuscript preparation, D.V. and E.C.S. All authors have read and agreed to the published version of the manuscript.

\subsubsection*{Funding}
This research received no external funding.


\subsubsection*{Acknowledgments}
The authors acknowledge the use of Miami University's RedHawk Cluster for the simulations.  E.C.S. acknowledges support by the Garland Professorship.

\subsubsection*{Conflicts of interest}
The authors declare no conflict of interest. 

\subsubsection*{Abbreviations}{
The following abbreviations are used in this manuscript:\\

\noindent 
\begin{tabular}{@{}ll}
BEC & Bose-Einstein condensate\\
STA & Shortcut-to-adiabaticity\\
CD & Counterdiabatic driving\\
CA & Constant acceleration\\
COM & Center-of-mass
\end{tabular}
}

\bibliographystyle{unsrt}
\bibliography{STAEntropy}

\end{document}